\documentstyle[preprint,prc,aps,epsfig]{revtex}
\begin{document}
\title{New Results in the Analysis of the $^{16}$O+$^{28}$Si Elastic Scattering
by Modifying the Optical Potential}
\author{I. Boztosun}
\address{Department of Physics, Erciyes University, 38039 Kayseri Turkey}
\date{\today}
\maketitle
\begin{abstract}
The elastic scattering of the $^{16}$O+$^{28}$Si system has been
analyzed with a modified potential within the framework of the
optical model over a wide energy range in the laboratory system
from 29.0 to 142.5 MeV. This system has been extensively studied
over the years and a number of serious problems has remained
unsolved: The explanation of the anomalous large angle scattering
data; the out-of-phase problem between theoretical predictions and
experimental data; the reproduction of the oscillatory structure
near the Coulomb barrier; the consistent description of angular
distributions together with the excitation functions data are just
some of these problems. We propose the use of a modified potential
method to explain these problems over this wide energy range. This
new method consistently improves the agreement with the
experimental data and achieves a major improvement on all the
previous Optical model calculations for this system.
\end{abstract}
%
%

{\bf Keywords:} $^{16}$O+$^{28}$Si Reaction, Optical model, elastic and inelastic
scattering, anomalous large angle scattering (ALAS), excitation
function.
%
\newpage
\section{Introduction}

The elastic and inelastic scattering between heavy ions have been
amongst the main sources of information about complex nuclei over
the last 40 years.\cite{Bra81,Bra82,Sci97}. There have been
numerous experimental investigations for the systems with a
combined mass number of $A_{P}+A_{T}\le60$,  and these
investigations have displayed a common unexpected feature near
$\theta_{CM}$=180$^{\circ}$ for the elastic and inelastic
scattering cross-sections.

The physical origin of the observed structure is not yet fully
understood \cite{Bra82,Boz99,Bozth,Bozth1,Bozth2,Boz1} and
presents a challenge to different approaches that have been
proposed to explain it. These approaches range from the occurrence
of possibly overlapping shape resonances \cite{Bar78} and the
scattering from surface-transparent Optical potentials
\cite{Kah79} to more exotic effects like explicit
parity-dependence of the ion-ion potential \cite{Deh78,Kub79}. The
first approach that has been more popular with researchers so far
attempts to describe the data by invoking the properties of the
average Optical potential
\cite{Kah79,Sat77,Ter78,Ren78,Kah83a,Mac80,Kob83,Kob84}. On the
other hand, in the second approach, the structure in the
excitation function is associated with isolated or nearly isolated
partial-wave resonances superimposed on the scattering properties
of a standard Optical potential \cite{Bra81,Bra82}. At present,
none of these approaches provides a consistent explanation for all
the existing data for this system.

Consequently, the following problems continue to exist for this
reaction \cite{Bozth,Boz1,Boz2,Bozlett}: $(1)$ The explanation of
anomalous large angle scattering data; $(2)$ the reproduction of
the oscillatory structure near the Coulomb barrier; $(3)$ the
out-of-phase problem between theoretical predictions and
experimental data; $(4)$ the consistent description of angular
distributions together with the excitation functions data.

Within the framework of the coupled-channels method, we have
analyzed such light heavy-ion reactions that pose the
above-mentioned problems \cite{Bozth,Boz1,Boz2,Bozlett}. The new
approach that we proposed within the coupled-channels method has
successfully explained the experimental data over wide energy
ranges for different systems such as $^{12}$C+$^{12}$C,
$^{16}$O+$^{28}$Si, $^{16}$O+$^{24}$Mg and $^{12}$C+$^{24}$Mg. One
feature we observed in these analyzes was that these reactions
were extremely sensitive to the shape of the nuclear potential in
the surface region. In this paper, by taking this feature into
account, we consider an extensive investigation of the elastic
scattering of this system at numerous energies by using a modified
potential. Similar to the previously conducted coupled-channels
analyzes for such systems \cite{Bozth,Boz1,Boz2,Bozlett}, this new
technique modifies the shape of the potential at the surface.
Thus, we aim to address the above-mentioned problems within the
framework of the Optical model and to obtain results that are as
good as the results of the coupled-channels method. Accordingly,
we analyze the experimental data from 29.0 to 142.5 MeV in the
laboratory system over the whole angular range up to
180$^{\circ}$. The 180$^{\circ}$ elastic scattering excitation
function has also been studied over this energy range. Extensively
modified version of the code {\it CHUCK} \cite{Kunz} has been used
for the all calculations.

In the next section, we introduce our Optical model and potential parameters to
explain the observed experimental data. Then, we show the results of
these analyzes in section \ref{results} from E$_{Lab}$=29.0 MeV to 142.5 MeV.
Our conclusion is given in Section \ref{conc}.
\section{The Model}

The standard Optical model with folding model potentials or with
similar phenomenological potentials such as the square of the
Woods-Saxon has failed to describe certain aspects of the
experimental data. Therefore, similar to Mackintosh, Kobos and
Satchler's work \cite{Mac80,Kob83,Kob84}, in the present
calculations, our total real potential consists of the nuclear
potential, $V_{Nuclear}$, with two small additional potentials,
$\left [ U(r)=U_{1}(r)+U_{2}(r)\right ]$:
\begin{equation}
V_{total}(r)= \underbrace{V_{Nuclear}(r)+ U(r)}_{Real \; potential}+
V_{Coulomb}(r)+ V_{Centrifugal}(r)
\end{equation}
The nuclear potential is assumed to have the square of a Woods-Saxon shape
and the parameters are fixed to reproduce the folding model potential of
Mackintosh, Kobos and Satchler \cite{Mac80,Kob83,Kob84}:
\begin{equation}
V_{Nuclear}(r) = \frac{-V_{0}}{(1+exp(r-R)/a)^{2}}.
\label{realx}
\end{equation}
where $V_{0}$=761.5 MeV and R=$r_{0}$($A_{a}$$^{1/3}$+$A_{A}$$^{1/3}$)
with $r_{0}$=0.75 fm  and a=1.425 fm. The parameters of the nuclear
potential are fixed as a function of energy and kept constant
in the present calculations although small changes were observed to improve
the quality of the fits.

The Coulomb potential \cite{Sat83} due to a charge $Z_{a}e$
interacting with a charge $Z_{A}e$, distributed uniformly over a
sphere of radius $R_{c}$, is also added.
\begin{eqnarray}
V_{Coulomb}(r) & = & \frac{1}{4\pi\epsilon_{\circ}}\frac{Z_{a}Z_{A}e^{2}}{r}, \hspace*{3cm} r\geq R_{c} \\
     & = & \frac{1}{4\pi\epsilon_{\circ}}\frac{Z_{a}Z_{A}e^{2}}{2R_{c}}(3-
\frac{r^{2}}{R_{c}^{2}}), \hspace*{1.3cm} r < R_{c}
\end{eqnarray}
where $R_{c}$=7.78 fm is the Coulomb radius, and $Z_{a}$ and
$Z_{A}$ denote the charges of the projectile $a$ and the target
nuclei $A$ respectively.

\begin{figure}[h]
\epsfxsize 12.5cm \centerline{\epsfbox{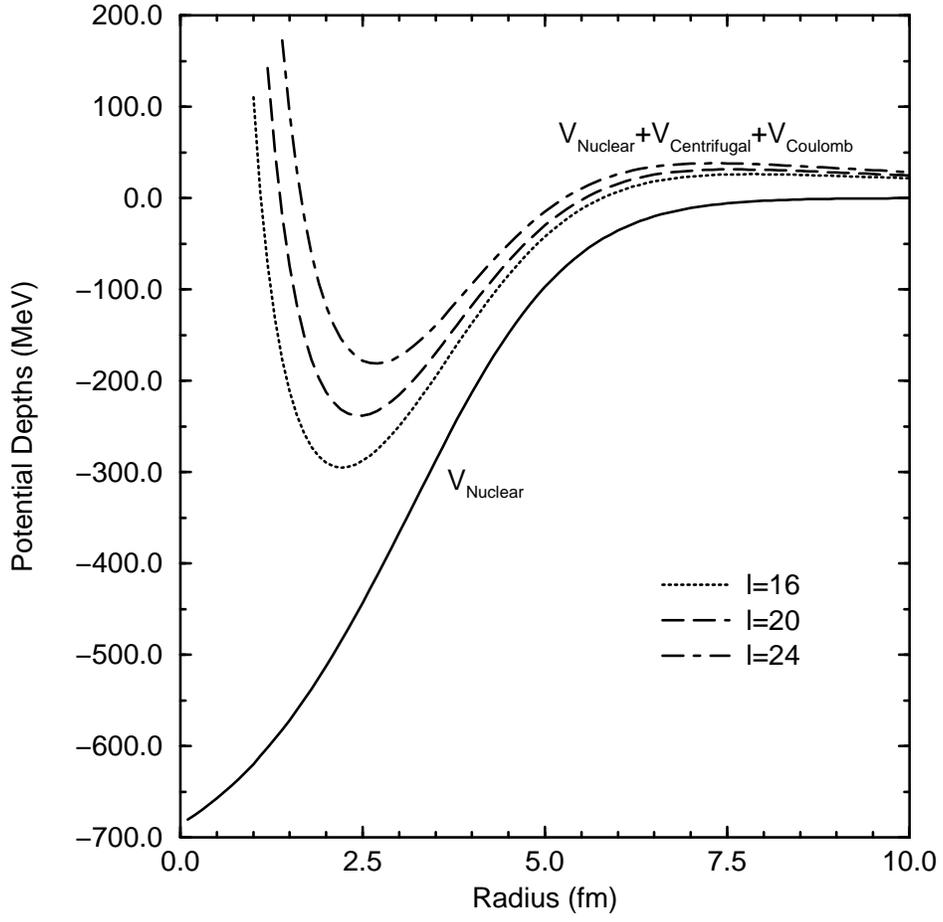}} \vskip+0.5cm
\caption{Interaction potential between $^{16}$O and $^{28}$Si is
plotted against the separation R for various values of the orbital
angular momentum quantum number, $l$. The parameters are given in
the text.} \label{fig:real}
\end{figure}

The sum of the nuclear, Coulomb and the centrifugal potentials is
shown in figure~\ref{fig:real} for various values of the orbital
angular momentum. The superposition of the attractive and
repulsive potentials results in the formation of a potential
pocket. The width and depth of the pocket depend on the orbital
angular momentum quantum number for a given nuclear potential.
This pocket is very important for the interference of the barrier
and internal waves, which creates the oscillatory structure
observed in the cross-section.

The significance of the two small additional potentials
should be emphasized here, since without them, it would be
impossible to fit the experimental data in the elastic
scattering calculations. These two small
additional potentials are the derivatives of the Woods-Saxon shape and
the parameters are shown in table \ref{small} where
\begin{equation}
U(r)=4U_{1}a_{1}df(r,R_{1},a_{1})/dr+4U_{2}a_{2}df(r,R_{2},a_{2})/dr
\label{u}
\end{equation}
\begin{equation}
f(r,R_{i},a_{i}) = \frac{1}{(1+exp((r-R_{i})/a_{i}))}
\end{equation}

\begin{figure}[ht]
\epsfxsize 13.5cm \centerline{\epsfbox{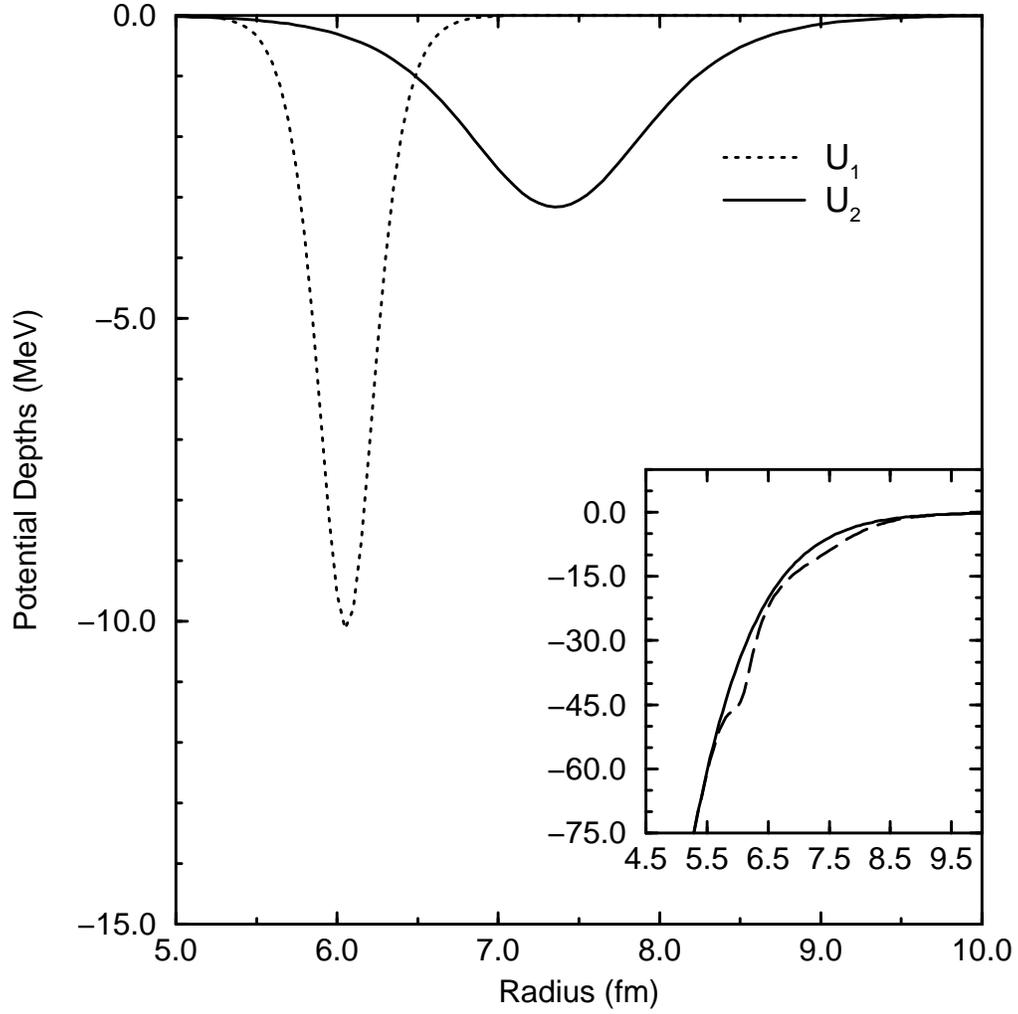}} \caption{The
shapes of two small additional potentials $U_{1}$ and $U_{2}$ are
displayed by dotted and solid lines respectively. The inserted
figure shows their effects on the nuclear potential with a
long-dashed line.} \label{fig:small}
\end{figure}

\begin{figure}[ht]
\epsfxsize 13.5cm \centerline{\epsfbox{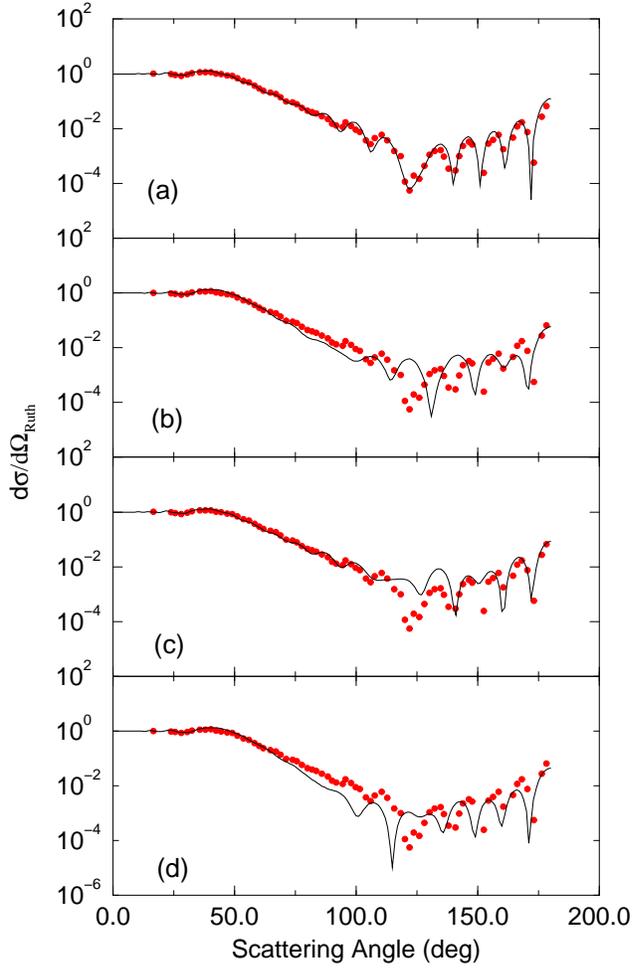}}
\caption{The effect of the two small additional potentials for the
$E_{Lab}$=41.17 MeV. (a) reflects that the best fit is obtained
with the inclusion of both potentials, (b) displays the situation
without the inclusion of $U_{2}$ and (c), without the inclusion of
$U_{1}$. Finally, (d) displays the outcome when both potentials
are ignored.} \label{fig:e262small}
\end{figure}

As shown in the insert of figure \ref{fig:small}, they create two
minima in the nuclear potential between $\sim$5.5 fm and $\sim$9.0
fm. The effect of these potentials can be understood in terms of
the interference between the internal and barrier waves that
correspond to a decomposition \cite{Lee78,Bri77} of the scattering
amplitude into two components, the inner and external waves. The
inner wave comes from the reflection at the inner face of the
total real potential pocket and the external wave comes from the
reflection at the outer barrier (see figure \ref{fig:real} for the
pocket in the total real potential). The presence of the two small
potentials affects the phases and magnitudes of these internal and
external components. We observe from the parameters in table
\ref{small} that the two small additional potentials are not
strong enough to produce pockets in the total real potential
although they have a very significant effect on the scattering.
This is demonstrated in figure \ref{fig:e262small} at 41.17 MeV.
In that figure, (d) shows the cross-sections obtained when both
potentials are omitted. As a result, the calculated oscillations
are out-of-phase with the measured ones even at the intermediate
angles. Without these potentials, we were unable to refit the data
by merely varying the parameters of the real and imaginary
potentials. Thus, the oscillatory structure could not be
reproduced correctly.

\begin{figure}[ht]
\epsfxsize 13.5cm \centerline{\epsfbox{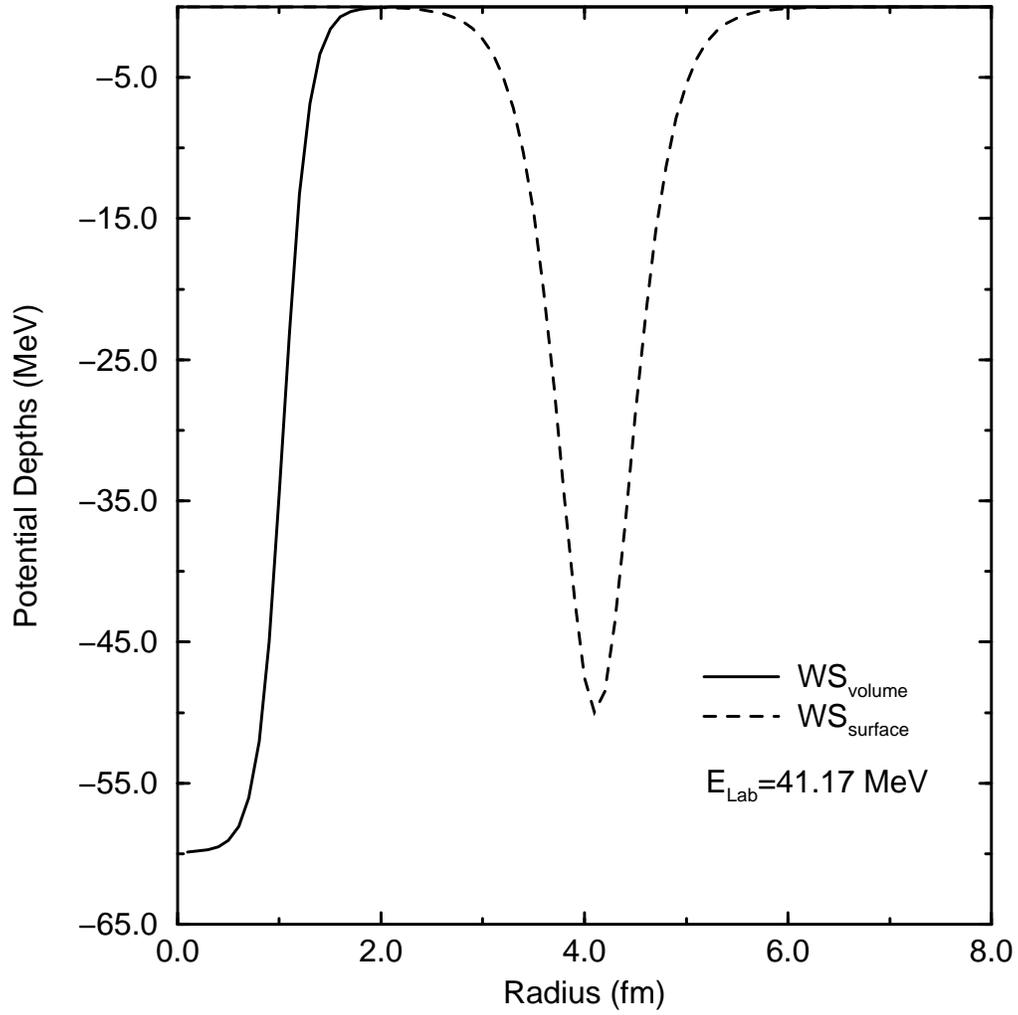}}
    \caption{The volume and the surface components of the imaginary
    potential at $E_{Lab}$=41.17 MeV for the $^{16}$O+$^{28}$Si system.}
\label{fig:imag}
\end{figure}

The imaginary part of the potential was taken
as the sum of a Woods-Saxon volume and the surface potential \cite{Kob84}:
\begin{equation}
W(r)=-W_{V}f(r,R_{V},a_{V})+4W_{S}a_{S}df(r,R_{S},a_{S})/dr
\end{equation}
\begin{equation}
f(r,R,a) = \frac{1}{(1+exp((r-R)/a))}
\end{equation}
with $W_{V}$=59.9 MeV, $a_{V}$=0.127 fm and $W_{S}$=50.0 MeV,
$a_{S}$=0.250 fm. These parameters were also fixed in the calculations
and only their radii were increased linearly with increasing energy
according to the following formulae.
\begin{equation}
R_{V} = 0.06084E_{CM}-0.544
\label{imagv}
\end{equation}
\begin{equation}
R_{S} = 0.2406E_{CM}-2.191
\label{imags}
\end{equation}
The imaginary potentials are shown in figure \ref{fig:imag} for
$E_{Lab}$=41.17 MeV.

Moreover, the relative significance of the volume
and surface components of the imaginary potential was evaluated for all
the energies. For higher energies, omitting the volume term predominantly
affected the amplitude of the cross-section at large angles. However,
this effect was small and negligible at lower energies.
Omitting the surface term increased the cross-sections at large angles which
were as much as two orders of magnitude. It was observed that
this term had a significant effect at all the considered energies.
\section{The Results}
\label{results}
\begin{figure}
\epsfxsize 13.5cm \centerline{\epsfbox{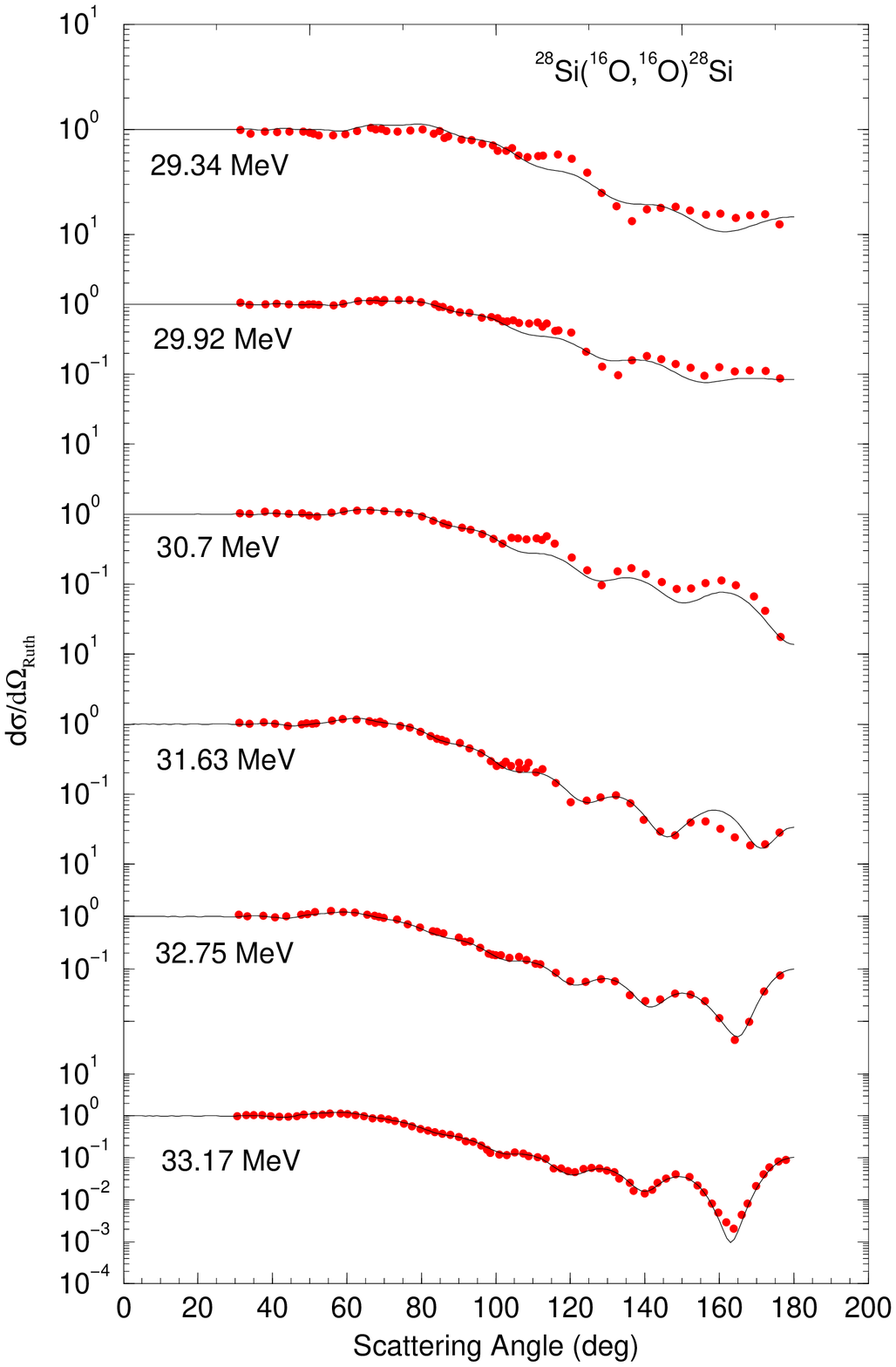}}
\caption{The elastic scattering results obtained by using the
single-channel Optical model calculations for the
$^{16}$O+$^{28}$Si system.} \label{elastic1}
\end{figure}

\begin{figure}
\vspace{22cm} \epsfxsize 13.5cm
\centerline{\epsfbox{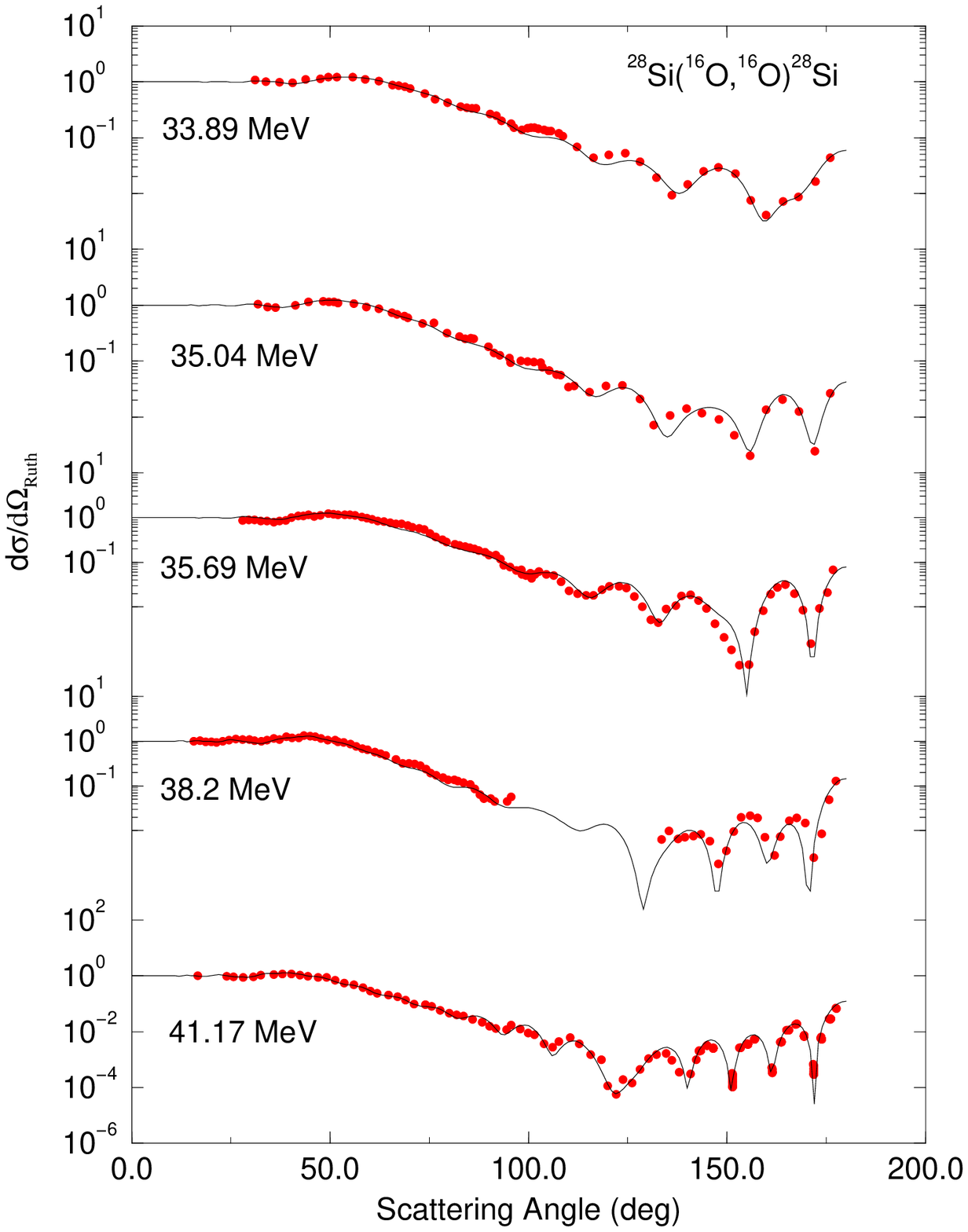}} \caption{Elastic
scattering results obtained by using the single-channel Optical
model calculations for the $^{16}$O+$^{28}$Si system ({\it
continued from figure \ref{elastic1}}).} \label{elastic2}
\end{figure}

\begin{figure}
\epsfxsize 13.5cm \centerline{\epsfbox{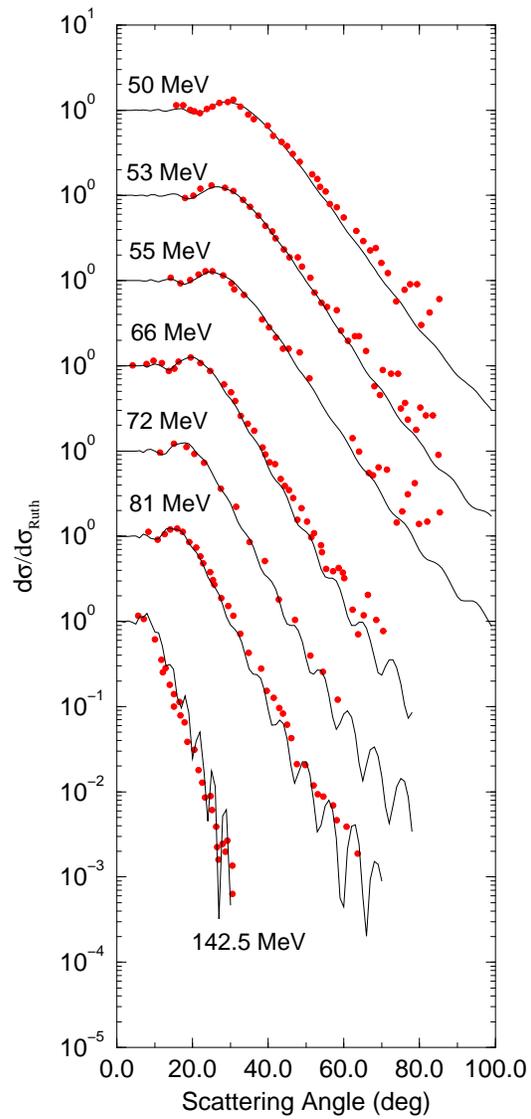}}
\caption{Results of the elastic scattering calculation for forward angles
at higher energies by using the single-channel Optical model
calculations.}
\label{elhigh}
\end{figure}

Using the above-described Optical model with fixed
real and linearly increasing imaginary potential parameters, we
have analyzed the experimental data from 29.0 MeV to 142.5 MeV in
the laboratory system. The comparisons between experimental data
and the Optical model fits are shown in figures \ref{elastic1},
\ref{elastic2} and \ref{elhigh}. The $\chi^{2}$ values for some of
the energies studied are also shown in table \ref{chi1}. As it can
be seen from these figures and the $\chi^{2}$ table, we have
obtained excellent agreement with the experimental data over the
whole energy range considered.

\begin{figure}
\epsfxsize 13.5cm \centerline{\epsfbox{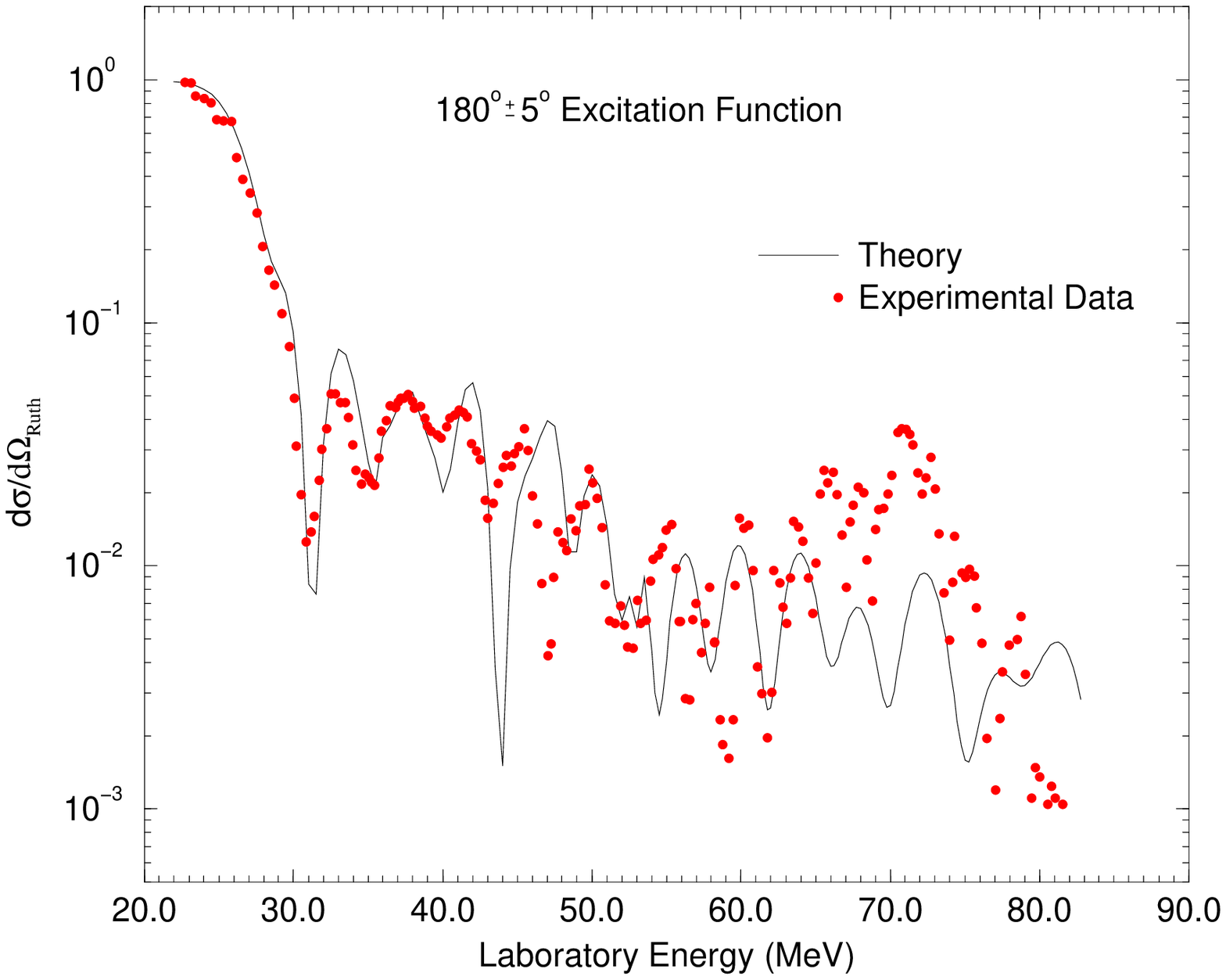}}
\caption{Comparison of the calculated 180$^{\circ}$ elastic scattering
excitation function with the measured experimental data. The theoretical calculation is averaged
over the last 5$^{\circ}$.} \label{exc}
\end{figure}

We have also analyzed the averaged value of the excitation
function over the angular range 180$^{\circ}$$\pm$5$^{\circ}$ with
these potentials. The radii of the imaginary potential increased
linearly with energy up to 54.0 MeV according to equations
(\ref{imagv}) and (\ref{imags}). Beyond this energy, we do not
have the available experimental data for the large angles.
Therefore, as shown in table \ref{imaghigh}, the radii were kept
constant at that energy and were not changed for the higher
energies. The result is compared with the measured values in
figure \ref{exc}. The overall agreement is quite good up to about
50 MeV where experimental data are available at large angles. In
particular, the maxima and minima in the excitation function are
obtained correctly. However, it should be emphasized that the
calculated excitation function is much more sensitive to the
details of the Optical potential than are the fits to the angular
distributions. For example, omitting two small potentials
completely destroys the agreement with the excitation function.

We were able to obtain almost the same agreement with the
experimental data above 50 MeV regardless of these two small
potentials. This shows that for higher energies, the
barrier/internal wave decomposition is not very important and thus
we may remove these two small potentials. By using the same
nuclear potential and the same type of imaginary potentials, but
by readjusting the depth and diffuseness of its surface component,
we were able to fit the experimental data from $E_{Lab}$=50 MeV to
142.5 MeV. These results are shown in figure \ref{elhigh} and the
relevant parameters are shown in table \ref{imaghigh}.
\section{Summary}
\label{conc}
We have shown a consistent description of the elastic scattering of
the $^{16}$O+$^{28}$Si system from 29.0
MeV to 142.5 MeV in the laboratory system by using the Optical model
calculations. In the introduction, we presented the problems that this
reaction manifests. We attempted to find a consistent solution to these problems.
However, within the standard Optical model calculations, we failed, as others did, to
describe certain aspects of the data.

Our previous elastic and inelastic analyzes with light-heavy ion
reactions using the coupled-channels formalism had already shown
that these reactions are extremely sensitive to the shape of the
potential in the surface region. By taking this feature into
account, we have used two small additional potentials that modify
the shape of the potential in the surface region. As a result, we
have obtained excellent agreement with the experimental data over
a wide energy range similar to that of the coupled-channels
method. The comparison of the results indicates that a global
solution to the problems relating to the scattering observables of
this reaction over a wide energy range has been provided by this
method. Finally, it should be noted that although these additional
two small potentials we used are very small and do not create a
pocket in the total nuclear potential, they are very effective for
the interference of the barrier and internal waves, which creates
the oscillatory structure observed in the cross-section. Further
work in order to derive this term from a microscopic viewpoint is
still under-progress.
\section*{Acknowledgments}
Author wishes to thank W.D.M. Rae and N. Ay\c{s}e Boztosun for
useful comments. He is also grateful to the members of the Nuclear
Physics Laboratory in Oxford University.

%
\begin{table}[h]
\begin{center}
\begin{tabular}{llllllllll}
 $U_{1}$ & $R_{1}$ & $a_{1}$ & $U_{2}$ & $R_{2}$ & $a_{2}$\\
 (MeV) & (fm) & (fm) & (MeV) & (fm) & (fm) & \\     \hline
  10.14 & 6.057 & 0.177 & 3.17 & 7.358 & 0.371\\
\end{tabular}
\end{center}
\caption{The parameters of the two small additional potentials.}
\label{small}
\end{table}
\begin{table}[h]
\begin{center}
\begin{tabular}{llllllllll}
$W_{V}$ & $R_{V}$ & $a_{V}$ & $W_{D}$ & $R_{D}$ & $a_{D}$\\
 (MeV) & (fm) & (fm) & (MeV) & (fm) & (fm) & \\     \hline
  59.90 & 1.3925 & 0.127 & 31.25 & 5.4697 & 0.550\\
\end{tabular}
\end{center}
\caption{The parameters of the potentials required to fit the higher energy data.
These are the values of the imaginary potential at $E_{Lab}$=50 MeV and they were
fixed to reproduce the structure for forward angles at high energies for $^{16}$O+$^{28}$Si.}
\label{imaghigh}
\end{table}
\begin{table}
\begin{center}
\begin{tabular}{ccccccc} 
& Laboratory Energy & Optical Model \\ \hline
& 29.34 & 1.2 \\
& 29.92 & 1.4 \\
& 30.70 & 1.9 \\
& 31.63 & 1.5 \\
& 32.75 & 0.3 \\
& 33.17 & 0.8 \\
& 33.89 & 0.9 \\
& 35.04 & 2.0 \\
& 35.69 & 8.0 \\
& 38.20 & 7.9  \\
& 41.17 & 16.5 \\
\end{tabular}
\end{center}
\caption{$\chi^{2}$ values for some of the energies studied.}
\label{chi1}
\end{table}


\begin{thebibliography}{99}
\bibitem{Bra81} P. Braun-Munzinger, G.M. Berkowitz, M. Gai, C.W. Jachcinski, T.R. Renner,
C.D. Uhlhorn, J. Barrette and M.J. LeVine, Phys. Rev. C {\bf 24} (1981) 1010.
\bibitem{Bra82} P. Braun-Munzinger and J. Barrette, Phys. Reports {\bf 87} (1982) 209.
\bibitem{Sci97} W. Sciani, A. Lepine-Szily, F.R. Lichtenthaeler, P. Fachini, L.C. Gomes,
G.F. Lima, M.M. Obuti, J.M. Jr Oliveira and A.C.C. Villari,  Nucl.
Phys. {\bf A620} (1997) 91.
\bibitem{Boz99} I. Boztosun and W.D.M. Rae,  Proceedings of the 7$^{th}$ International
Conference on Clustering Aspects of Nuclear Structure and
Dynamics, Edited by M. Korolija, Z. Basrak and R. Caplar,
World-Scientific-2000 (143).
\bibitem{Bozth} I. Boztosun, DPhil Thesis, Oxford University, 2000.
\bibitem{Bozth1} I. Boztosun and W.D.M. Rae, Phys. Rev. C {\bf 63} (2001) 054607.
\bibitem{Bozth2} I. Boztosun, Phys. At. Nuclei {\bf 65} (2002) 607.
\bibitem{Boz1} I. Boztosun and W.D.M. Rae, Phys. Rev. C {\bf 65} (2002) 024603.
\bibitem{Bar78} J. Barrette, M.J. LeVine, P. Braun-Munzinger, G.M. Berkowitz, M. Gai,
J.M. Harris and C.W. Jachcinski, Phys. Rev. Lett. {\bf 40} (1978) 445.
\bibitem{Kah79} S. Kahana, B.T. Kim and M. Mermaz, Phys. Rev. C {\bf 20} (1979) 2124.
\bibitem{Deh78} D. Dehnhard, V. Shkolnik and M.A. Franey , Phys. Rev. Lett. {\bf 40} (1978) 1549.
\bibitem{Kub79} S. Kubono, P.D. Bond and C.E. Thorn, Phys. Lett. {\bf 81B} (1979) 140.
\bibitem{Sat77} G.R. Satchler, Nucl. Phys. {\bf A279} (1977) 493.
\bibitem{Ter78} K.O. Terenetski and J. D. Garrett, Phys. Rev. C {\bf 18} (1978) 1944.
\bibitem{Ren78} T.R. Renner, J. P. Schiffer, D. Horn, G. C. Ball and W. G. Davies, Phys. Rev. C {\bf 18} (1978) 1927.
\bibitem{Kah83a} S. Kahana, J. Barrette, B. Berthier, E. Chavez, A. Greiner and
M. C. Mermaz, Phys. Rev. C {\bf 28} (1983) 1393.
\bibitem{Mac80} R.S. Mackintosh and A.M. Kobos, Phys. Rev. C {\bf 26} (1982) 1766.
\bibitem{Kob83} A.M. Kobos, G.R. Satchler and R.S. Mackintosh, Nucl. Phys. {\bf A395} (1983) 248.
\bibitem{Kob84} A.M. Kobos and G.R. Satchler, Nucl. Phys. {\bf A427} (1984) 589.
\bibitem{Boz2} I. Boztosun and W.D.M. Rae, Phys. Rev. C {\bf 64} (2001) 054607.
\bibitem{Bozlett} I. Boztosun and W.D.M. Rae, Phys. Lett. {\bf 518B} (2001) 229.
\bibitem{Kunz} P.D. Kunz, CHUCK, a coupled-channels code, unpublished.
\bibitem{Sat83} G.R. Satchler, {\tt Direct Nuclear Reactions} (Oxford University Press, Oxford 1983) and
{\tt Introduction to Nuclear Reactions} (The Macmillan Press Ltd,
London 1980).
\bibitem{Lee78} S.Y. Lee, Nucl. Phys. {\bf A311} (1978) 518.
\bibitem{Bri77} D.M. Brink and N. Takigawa, Nucl. Phys. {\bf A279} (1977) 159.
\end{thebibliography}
\end{document}